\documentclass{aa}  

\usepackage{graphicx}
\usepackage{txfonts,textcomp}
\usepackage{lineno}

\usepackage{natbib,twoopt}
\bibpunct{(}{)}{;}{a}{}{,}             

\usepackage{hyperref}
\hypersetup{
colorlinks=true,    
linkcolor=red,      
citecolor=blue,     
filecolor=magenta,  
urlcolor=black      
}

\newcommand{\swift}{\emph{Swift}}

\begin{document} 

\title{Highly luminous supernovae associated with gamma-ray
  bursts\\\Large II. The luminous blue bump in the afterglow of GRB 140506A}

\author{D.~A.~Kann\thanks{Deceased}
\inst{1,2}
\and
A.~Rossi\thanks{E-mail: andrea.rossi@inaf.it}
\inst{3}
\and
S.~R.~Oates
\inst{4}
\and
S.~Klose
\inst{5}
\and
M.~Blazek 
\inst{6}
\and
J.~F.~Ag\"u\'i Fern\'andez
\inst{2,6}
\and
A.~de Ugarte Postigo
\inst{7,8}
\and
C.~C.~Th\"one
\inst{9}
\and 
S. Schulze
\inst{10}
}

\institute{Hessian Research Cluster ELEMENTS, Giersch Science Center, Max-von-Laue-Stra$\beta$e 12, Goethe University Frankfurt, Campus Riedberg, D-60438 Frankfurt am Main, Germany 
\and
Instituto de Astrof\'isica de Andaluc\'ia (IAA-CSIC), Glorieta de la Astronom\'ia s/n, 18008 Granada, Spain  
\and
INAF - Osservatorio di Astrofisica e Scienza dello Spazio, via Piero Gobetti 93/3, 40129 Bologna, Italy 
\and
Department of Physics, Lancaster University, Lancs LA1 4YB, UK 
\and
Th\"uringer Landessternwarte Tautenburg, Sternwarte 5, 07778 Tautenburg, Germany 
\and
Centro Astron\'omico Hispano en Andaluc\'ia, Observatorio de Calar Alto, Sierra de los Filabres, 04550 G\'ergal, Almer\'ia, Spain 
\and 
Artemis, Universit\'{e} de la C\^ote d'Azur, Observatoire de la C\^ote d'Azur, CNRS, 06304 Nice, France 
\and 
Aix Marseille Univ, CNRS, LAM Marseille, France
\and
Astronomical Institute, Czech Academy of Sciences, Fri\v cova 298, Ond\v rejov, Czech Republic 
\and
Center for Interdisciplinary Exploration and Research in Astrophysics (CIERA), Northwestern University, 1800 Sherman Ave., Evanston, IL 60201, USA 
}

\date{Received 30 September 2021 / Accepted 30 January 2024}

\authorrunning{Kann et al.}
\titlerunning{The Luminous GRB-SN 140506A}

\abstract
%
{The supernovae (SNe) associated with gamma-ray bursts (GRBs) are
generally seen as a homogeneous population, but at least one exception
exists: the highly luminous SN 2011kl associated with
the ultra-long GRB 111209A. Such outliers may also exist for more
typical GRBs.}
%
{Within the context of a systematic analysis of photometric signatures
of GRB-associated SNe, we found an anomalous bump in the late-time
transient following GRB 140506A at redshift $z$=0.889. We
hereby aim to show this bump is significantly more luminous and blue
than usual SNe following GRBs.}
%
{We compiled all available data from the literature and added a full
analysis of the \emph{Swift}/UVOT data, which allowed us to trace the
light curve from the first minutes all the way to the host galaxy and to construct a broad spectral energy distribution (SED) of the afterglow that extends the
previous SED analysis based on ground-based spectroscopy.}
%
{We find robust evidence of a
late-time bump following the afterglow that shows evidence
of a strong color change, with the spectral slope becoming flatter
in the blue region of the spectrum. This bump can be interpreted as a
luminous SN bump that is spectrally dissimilar to typical
GRB-SNe. Correcting it for the large line-of-sight extinction
makes the SN associated with GRB 140506A the most luminous detected so far. Even so, it would be in agreement with a luminosity-duration
relation of GRB-SNe.}
%
{While not supported by spectroscopic evidence, it is likely the 
bump following GRB 140506A is the signature of an SN that is
spectrally dissimilar to classical GRB-SNe and more similar to SN
2011kl -- while being associated with an average GRB, indicating the
GRB-SN population is more diverse than previously thought and can reach luminosities comparable to those of superluminous SNe.}


\maketitle

\section{Introduction}

The link between ``long gamma-ray bursts'' (GRBs) usually exhibiting a duration of $>2$
s \citep[however, see][]{Ahumada2021NatAs,Zhang2021NatAs,Rossi2021ApJ},
also labeled ``Type II GRBs'' in a more physically motivated,
duration-independent classification scheme
\citep{Zhang2009ApJ,Kann2011ApJ}, and the explosive deaths of massive
stars has now been firmly established (see
\citealt{Woosley2006ARAA,Cano2017AdAst} for reviews). Beginning with
the prototypical, well-studied supernova (SN) 1998bw associated with the
low-luminosity, nearby GRB 980425
\citep{Galama1998Nature,Clocchiatti2011AJ}, all GRB-SNe that have been
studied well have been found to be explosions of highly stripped stars
(so-called broad-lined Type Ic SNe), which are missing signatures of hydrogen
and helium in their spectra. Perhaps surprisingly, these SNe have been
shown to be very similar in terms of luminosity, ejecta
and nickel masses, and energy release \citep{Melandri2014AA}, whether
they have been associated with a spectrally soft, low-luminosity X-ray
flash such as XRF 060218
\citep[e.g.][]{Pian2006Nature,Mazzali2006Nature}, a moderately soft
and energetic GRB such as GRB 030329
\citep[e.g.][]{Hjorth2003Nature,Matheson2003ApJ,Stanek2003ApJ}, or a
highly luminous and spectrally hard ``true cosmological'' GRB such as GRB
130427A \citep{Xu2013ApJ,Melandri2014AA}.

A decade ago, a startling outlier was discovered. GRB 111209A
was a GRB of extreme duration in gamma-rays
\citep{Golenetskii2011GCN,Gendre2013ApJ,Stratta2013ApJ}, leading,
together with the ``Christmas Burst'' GRB 101225A
\citep{Thoene2011Nature}, to the establishment of the ultra-long-duration GRB class \citep{Levan2014ApJ}. GRB 111209A was found to be
associated with a well-detected SN, SN 2011kl, which exhibited
properties very different from usual GRB-SNe, being both more
luminous, bluer, and spectrally dissimilar to typical Type Ic GRB-SNe,
and closer in resemblance, spectroscopically, to superluminous SNe
\citep[SLSNe][]{Greiner2015Nat,Mazzali2016MNRAS}. Notwithstanding its
extreme duration, both GRB 111209A and its afterglow are not
unprecedented \citep{Kann2018AA}, but a detailed study shows SN 2011kl
differing from GRB-SNe in many aspects \citep{Kann2019AA}.

This discovery immediately opened multiple new lines of inquiry. We now question whether all ultra-long GRBs are associated with anomalous GRB-SNe, and
  if so, whether they are similar to SN 2011kl or outliers in other aspects. Moreover, we would like to know if such peculiar, highly luminous GRB-SNe are exclusively
  associated with ultra-long GRBs or if they can also occur following
  ``standard'' GRBs.

So far, the lack of more detailed studies of ultra-long GRBs (which
are very rare) has prevented us from exploring the first question
further. However, here we present evidence of a highly
luminous GRB-SN associated with a ``standard'' GRB, which indicates
that these events are not limited to ultra-long GRBs.
In the context of a systematic study of late-time emission in GRB
afterglows at $z\lesssim1$, we
find that the per se unremarkable (at high energies) GRB 140506A,
which \emph{\emph{has}} been studied in detail stemming from its peculiar
line-of-sight in terms of extinction and spectral features
\citep{Fynbo2014AA,Heintz2017AA}, exhibits a bump at late times that
shows a strong color change from very red to very blue. This bump is
significantly brighter than the final host-galaxy magnitudes, making
the interpretation as a late-time SN component compelling. Furthermore, we analyzed spectral energy distribution (SED) of the afterglow, determined the dust extinction
along the line of sight, and placed the potential SN in a
larger context of extinction-corrected GRB-SNe.

The paper is organized as follows. In Sect.~\ref{obs}, we present
GRB 140506A and our observations of the afterglow.  We discuss
our analysis and our results in Sect.~\ref{results}, and we place the
event in context in Sect.~\ref{discussion} before concluding in
Sect. \ref{conclusions}.

In our calculations, we assume a flat Universe with a matter density
$\Omega_M=0.27$, a cosmological constant $\Omega_\Lambda=0.73$, and a
Hubble constant $H_0=71$ km s$^{-1}$ Mpc$^{-1}$
\citep{Spergel2003ApJS}, to remain in agreement with our older sample
papers. Errors are given at the $1\sigma$ level, and upper limits are given at the
$3\sigma$ level for a parameter of interest. For temporal and spectral
power laws, we assume $F(t,\nu)\propto t^{-\alpha}\nu^{-\beta}$;
therefore, temporally decaying light curves and spectra rising in
brightness toward the red (the typical situation for GRB afterglows)
have positive $\alpha$ and $\beta$ values.

\section{Observations of GRB 140506A \label{obs}}

\subsection{Prompt emission phase}

GRB 140506A triggered the \emph{Neil Gehrels Swift Observatory}
(\emph{Swift} hereafter, \citealt{Gehrels2004ApJ}) at 21:07:36 UT on 6
May, 2014 \citep{Gompertz2014GCN16214}. The satellite slewed
immediately, localizing the event precisely with X-ray, ultra-violet (UV), and optical
detections. It was a moderately bright GRB consisting of an initial
spike followed by several fainter emission episodes, and it also
triggered \emph{Fermi}/GBM \citep{Jenke2014GCN16220} and
Konus-\emph{Wind} \citep{Golenetskii2014GCN16223}. The
$T_{90}$\footnote{The time span over which 90\% of the integrated
  counts are emitted, beginning 5\% after the start of detection and
  ending at 95\%. This is a general duration measure used for GRBs.}
derived by \emph{Swift} was $111.1\pm9.6$ s
\citep{Markwardt2014GCN16218}, so clearly not of ultra-long
duration\footnote{$T_{90}$ from Konus-\emph{Wind} and \emph{Fermi}/GBM
  are about half this value.}. \cite{Tsvetkova2017ApJ} reported a
detailed analysis of the Konus-\emph{Wind} detection of GRB 140506A,
deriving a fluence in the $10-10000$ keV energy band of
$5.74^{+0.52}_{-0.34}\times10^{-6}$ erg cm$^{-2}$, and a spectrum that
is best fit by a cut-off power law with $\alpha_{\rm
  prompt}=1.32^{+0.23}_{-0.26}$ and $E_{\rm peak}=200^{+90}_{-42}$
keV. Using these parameters and the redshift $z=0.88911$
\citep{Fynbo2014AA}, we deduce \citep[following][]{Agui2021MNRAS} an
isotropic energy release in the rest-frame 1 keV to 100 MeV band of
$\log E_{\rm iso}/\textnormal{erg}=52.15\pm0.03$, implying GRB 140506A
is an average GRB that is neither particularly sub- nor super-luminous.

Nonetheless, the GRB features one interesting aspect, namely a
peculiar, red, and strongly curved afterglow spectrum, leading to a
detailed spectroscopic and photometric analysis by \cite{Fynbo2014AA},
which was followed by a host-galaxy study \citep{Heintz2017AA}. In the
late-time data presented in \cite{Fynbo2014AA}, the afterglow decay
flattened considerably, seemingly implying the host-galaxy level had
been reached; however, \cite{Heintz2017AA} showed this to be incorrect,
finding further significant decay between $\approx2$ months and a year
post-burst. They even remarked that the earlier plateau phase is similar
to what one would expect from a GRB-SN, but they did not pursue the topic
further.

\subsection{Follow-up observations}

\cite{Fynbo2014AA} presented ground-based photometric observations
starting 0.33 d after the GRB, with detections spanning from the GROND
$g^\prime$ band to the GROND $K_s$ band. Their last detection epoch is
68~d post-burst. They also obtained \emph{Very Large Telescope} X-shooter spectra at
  8.8 and 33~h after the burst and a spectrum at 52 days using the
  \emph{Magellan} telescope. \cite{Heintz2017AA} added host-galaxy data
spanning from $u^\prime$ to the \emph{Spitzer} IRAC1 band (at
$3.6\,\mu$m) taken about a year after the GRB, when any transient is
expected to have faded. 

To expand this data set both temporally and spectrally, we
analyzed the \emph{Swift} UltraViolet-Optical Telescope
\citep[UVOT,][]{Roming2005SSRv} data. UVOT began settled observations
of the field of GRB 140506A 108 s after the \emph{Swift} Burst Alert Telescope (BAT) trigger,
with initial results reported in \cite{Siegel2014GCN16219}. The data
(in AB mags) are given in Table \ref{obslog}. The last
low-significance detection is at 1.8 d, with upper limits until 30
d. We note that \cite{Fynbo2014AA} also analyzed the UVOT data and
only claimed detections for $ubv\,white$, with a low-S/N detection in
$uvw1$. While the S/N is certainly low ($<2\sigma$), we also report multiple
early detections in $uvm2$ and $uvw2$.

Before extracting count rates from the event lists, the astrometry was
refined following the methodology of \cite{Oates09MNRAS}. The source
counts were initially extracted using a source region with a 5\arcsec{}
radius. When the count rate dropped to below 0.5 counts per second, we used a
source region with a 3\arcsec{} radius. In order to be consistent with the
UVOT calibration, these count rates were then corrected to 5\arcsec{}
using the curve of growth contained in the calibration
files. Background counts were extracted using three circular regions
with radii of 10\arcsec{} located in source-free regions. The count rates
were obtained from the event and image lists using the \emph{Swift}
tools \texttt{uvotevtlc} and \texttt{uvotsource}, respectively. They
were converted to magnitudes using the UVOT photometric zero points
\citep{Poole2008MNRAS,Breeveld2011AIPC}. To improve the
signal-to-noise ratio (S/N), the count rates in each filter were binned
using $\Delta t/t=0.1$, leading to longer but deeper exposures at
later times. The early event-mode $white$ and $u$ finding-charts were
bright enough to be split into multiple exposures.

\section{Results \label{results}}
\subsection{The afterglow \label{AG}}

\begin{figure*}[t!]
\begin{center}
\centering 
\includegraphics[width=1.0\textwidth]{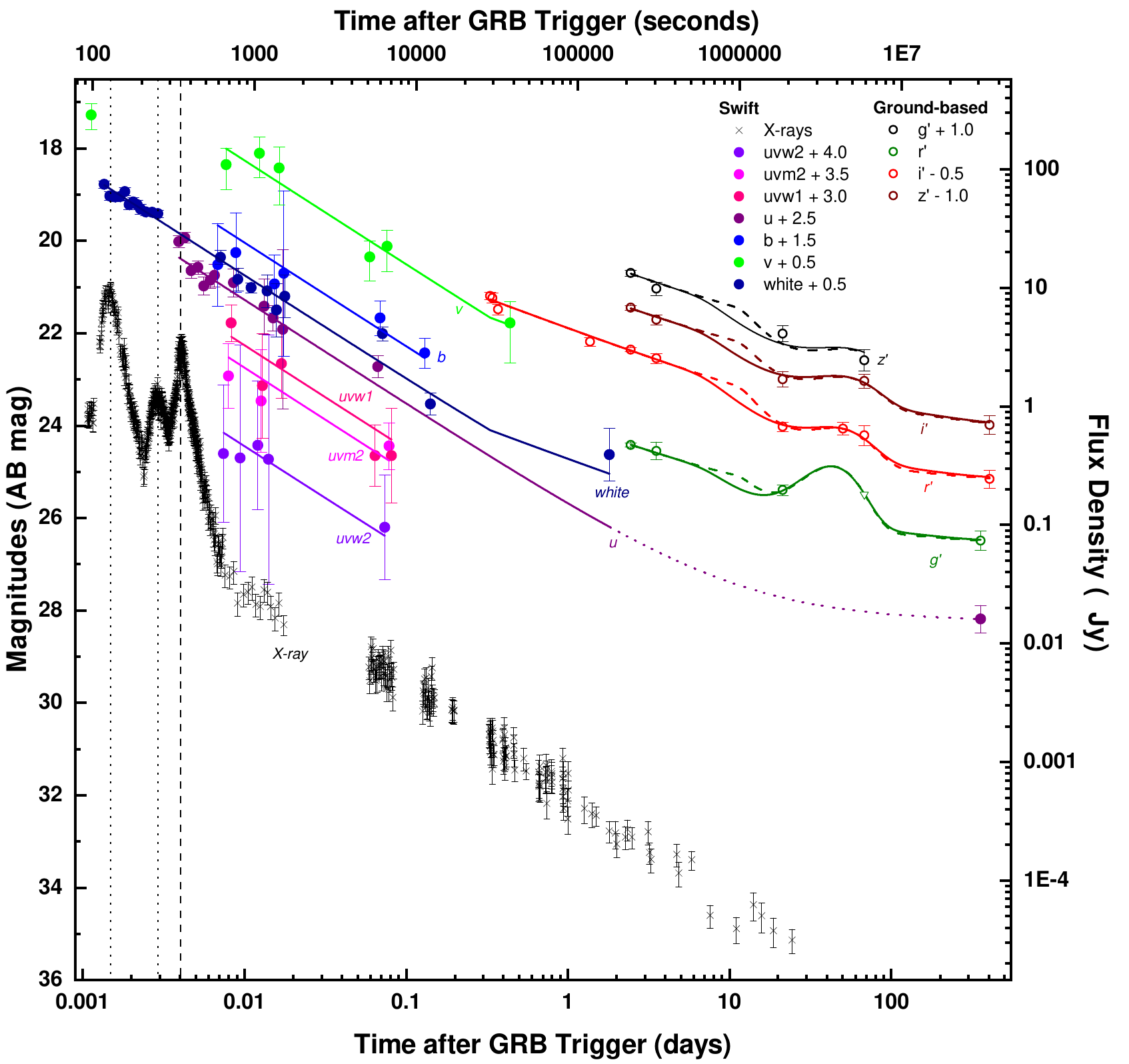} 
\caption{
  Observations of optical transient that followed
  GRB 140506A. UVOT data ($uvw2, uvm2, uvw1, u, b, v, white$) are
  presented in this paper (Table~\ref{obslog}). Ground-based $u^\prime
  g^\prime r^\prime i^\prime z^\prime$ magnitudes are from
 \cite{Fynbo2014AA} and 
  \cite{Heintz2017AA}. The downward-pointing triangle in $g^\prime$ at 68 days is
    an upper limit. We do not show UVOT
  upper limits and $JHK_s$ data as they add no relevant information.
  Data are corrected for Galactic extinction, given in the AB
  magnitude system, and additionally offset by the values given in the
  legend to improve legibility. The flux density scale is only valid
  for the (unshifted) $r^\prime$ band. The rest was shifted for
  clarity. The scale of the X-ray data was chosen arbitrarily. We show
  two possible fits with breaks at 7 (solid fit curve) and at 12
  (dotted fit curve) days,
  which include both the afterglow and the
  SN/host contributions.
  Dotted vertical lines mark the peaks of two
  early X-ray flares that have no counterparts in the UV/optical
  range, whereas the flare marked by the dashed vertical line is also
  detected by UVOT. The flatter decay after 0.33~d in the
  late UVOT $v$ and $white$ light curves is simply the shifted
  $r^\prime$-band fit (see text for more details). }
\label{lc}
\end{center}
\end{figure*}

The multicolor light curve of the optical transient (OT) that
followed GRB~140506A is shown in Fig.~\ref{lc}. UVOT data are
presented in Table~\ref{obslog}.
We did not add UVOT upper limits for the sake of legibility, and near-infrared (NIR)
$JHK_s$-band data \citep[from][]{Fynbo2014AA} as they do not
contribute notably to the light-curve fit. They are used in the SED,
however (Sect.~\ref{SED}). The early UVOT data, despite showing (for
the most part) low S/N detections with large errors, agree well with a
single power-law fit over all bands and the complete time span. The
one exception is the first two data points derived from the $u$ event-mode finding-chart at $\approx340-370$~s; these lie
$\sim3\sigma$ above the rest of the $u$-band data. The
\emph{Swift}/X-ray telescope (XRT, \citealt{Burrows2005SSRv}) data at
this time, as
given\footnote{\url{https://www.swift.ac.uk/xrt_curves/00598284/}} on
the XRT repository \citep{Evans2007AA,Evans2009MNRAS}, indeed show a
strong flare taking place right at this time, linking the elevated
emission to an additional component, likely stemming from internal
shocks. However, an even more powerful earlier X-ray flare, peaking at
$\approx120$~s, is not visible in the time-resolved $white$ finding
chart (Fig.~\ref{lc}). We find that the entire UVOT data are well-fit
($\chi^2/$d.o.f.=0.65) by a simultaneous, multiband, single-power-law
fit with decay of $\alpha_1=0.90\pm0.03$. This assumes that the evolution
of the afterglow is achromatic. While the early time prompt-emission flare may have a different SED, it only affects two data
points, and the X-ray light curve shows no further evidence of
flaring. The fit equation determines the afterglow magnitude in each
band (all data have been corrected for Galactic extinction following
\citealt{Schlafly2011ApJ}), which yields the UVOT-data SED.

The reddened and relatively faint afterglow implies that it is hardly
detected beyond $\approx0.14$ d, and therefore there is no useful
overlap with the ground-based data set from \cite{Fynbo2014AA}. We
therefore fit that data independently. It clearly shows a shallower
decay than before, which is an unusual but not unprecedented phenomenon. The
data extending from 0.33 to 3.5 days were again fit by single decay
for which we find $\alpha_2=0.54\pm0.03$. A large data
gap follows and then two further epochs (as well as a single unfiltered point,
calibrated to $r^\prime$, in-between) at 21 and 68 days (as well as the
host-galaxy observations about a year after the GRB). A simple
extrapolation of the $\alpha_2=0.54$ decay showed that at 21 days, the
fit strongly overestimated the actual data; therefore, a break must
have occurred, likely a jet break. However, as the following
plateau is clearly not the host-galaxy level, this would imply
either a strong, long-lasting flattening of the afterglow, which is
very unlikely, or a new emission component.

\subsection{The supernova \label{SNcomp}}

Despite the high redshift and the large line-of-sight extinction in
the GRB host galaxy, the most natural explanation for the
observed re-brightening of the light curve in the optical bands is
an SN following GRB 140506A, as \cite{Heintz2017AA} already speculated.

The SN components associated with GRBs were first studied
  systematically by \citet{Zeh2004ApJ}, who introduced the $k,s$
  context as described below. Based on the detailed light
  curves of the prototypical GRB-SN 1998bw at $z$=0.0085
  \citep{Galama1998Nature,Clocchiatti2011AJ}, and assuming that the
  general shape of GRB-SN light curves is identical, templates of SN
  1998bw can be created at other redshifts. These templates can then
be altered based on the luminosity factor $k$, with $k=1$ implying the
GRB-SN is just as luminous as SN 1998bw would be at the specific GRB
redshift in the specific rest-frame band corresponding to the
observer-frame band the measurements were taken in. Furthermore,
without changing its fundamental shape, the light curve's temporal
evolution can be compressed ($s<1$) or stretched ($s>1$) with the
stretch factor, $s$. This model fits almost all GRB-SN light curves
very well, 
and in particular that of SN
2011kl, despite it being over-luminous compared to SN 1998bw
\citep{Kann2019AA}.

We derived template light curves for SN 1998bw at the redshift of GRB
140506A, $z=0.88911$. At this redshift, the observer-frame $g^\prime
r^\prime i^\prime z^\prime$ bands lie between $uvm2$ and $uvw1$,
between $uvw1$ and $U$, between $U$ and $B$, and at about $g^\prime$
in the rest frame, respectively. SN 1998bw was observed densely in
$BVR_CI_C$ and with less follow-up in $U$. To emulate bands that lie
blueward of $U$ in the rest frame, a simple power law with
$F(\nu)\propto\nu^{-3}$ was assumed 
  (\citealt{Zeh2004ApJ}).
Therefore, we caution here
that results derived from the $g^\prime$ band come with a caveat of
uncertainty. As the data are sparse, with only $r^\prime$ having three
data points during the SN-dominated epochs, we fixed $s$ to be a
shared parameter between bands, and then we performed a simultaneous
afterglow+SN+host fit, where the afterglow evolved achromatically but
the SN can have different $k$ values for each band, and the
host-galaxy magnitudes were also individual for each band (for
$z^\prime$, we estimated a host magnitude $z^\prime=24.0$ mag based on
the SED plot of \citealt{Heintz2017AA}).

The large data gap from three to 21 days implies that the break time and
post-break decay slope are degenerate and, therefore, could not both be left as free parameters in the fit.  However, the X-ray light
curve deviates downward at $\approx6-7$ days, which would favor an
early and achromatic break time. Therefore, in the following we
assume the break to be at 7 d. The fit is good ($\chi^2/d.o.f.=0.76$),
and the resulting $k,s$ factors are shown in Table~\ref{SNFits}. The
fit is shown in Fig.~\ref{lc}.

\begin{figure}[t]
\includegraphics[width=1.0\columnwidth]{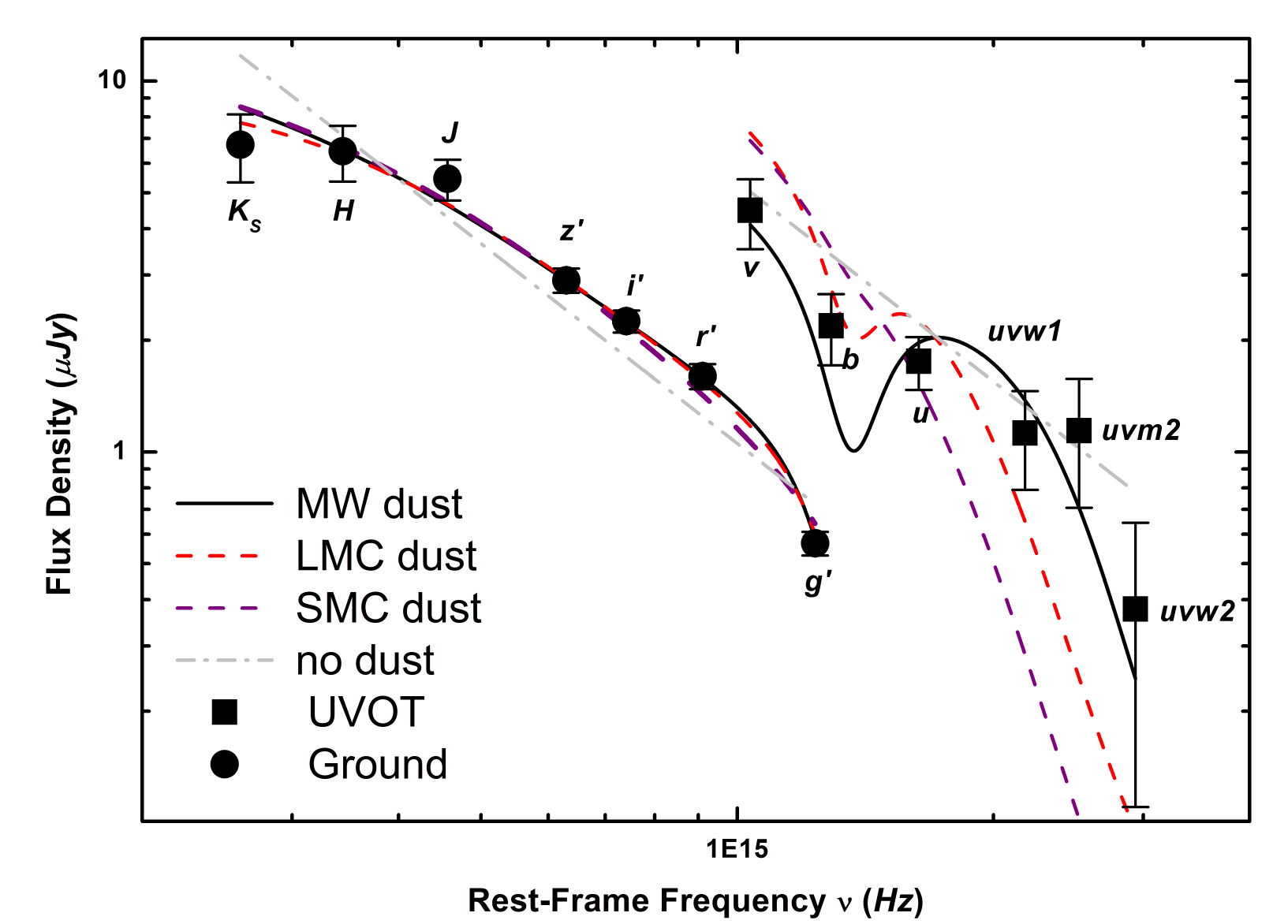}
  \caption{SED of the OT that
      followed GRB 140506A. The SED in the $v$-to-$uvw2$
      filter bands was derived from UVOT data at 24~h
    post-burst. The one covering the $K_s$-to-$g^\prime$
      filters is based on ground-based data at 7 d after the GRB. We
    show fits with no dust (gray dash-dot-dotted line), Milky Way dust
    (black compact line), Large Magellanic Cloud dust (red dashed
    line), and Small Magellanic Cloud dust (violet dashed line). The
    Milky Way dust fit is the best (see text for more details). }
\label{SEDFig}
\end{figure}

\begin{table}[t]
\caption{SED fit results for OT that followed GRB~140506A.}
\label{SEDFits} 
\centering                        
\begin{tabular}{c c r c}
\hline\hline   
$\beta$ & $A_{\rm V}$ (mag) & dust model & $\chi^2/$d.o.f. \\\hline
$1.79\pm0.07$ & --- & N/A & 6.09 \\
$0.24\pm0.27$ & $1.12\pm0.18$ & MW & 1.04 \\
$0.25\pm0.36$ & $1.50\pm0.25$ & LMC & 1.65 \\
$0.23\pm0.34$ & $1.11\pm0.22$ & SMC & 4.21 \\ 
\hline \hline
\end{tabular}
\end{table}

\begin{table*}[t]
\caption{Fit results for GRB 140506A supernova component.}
\label{SNFits} 
\centering                        
\begin{tabular}{l c c l}
\hline\hline   
Fit model & $s$ & $k$ (fitted) & $k$ (extinction-corrected) \\\hline
7 day break & $1.96\pm0.11$ & $k_{g^\prime}=5.45\pm1.21$ & $k_{g^\prime}=76.96\pm17.22$ \\
 &  & $k_{r^\prime}=1.19\pm0.39$ & $k_{r^\prime}=6.65\pm2.33$ \\
 &  & $k_{i^\prime}=1.20\pm0.39$ & $k_{i^\prime}=4.95\pm1.75$ \\
 &  & $k_{z^\prime}=1.00\pm0.41$ & $k_{z^\prime}=3.32\pm1.45$ \\
\hline \hline
\end{tabular}
\end{table*}

\subsection{Spectral energy distribution \label{SED} }

The equation used to fit the afterglow+SN+host uses overlapping
components; therefore, the normalizations that define the afterglow
magnitudes at 1 / 7 d are pure afterglow values and directly represent
the SEDs at these times. The afterglow fits are
achromatic, implying that these SEDs are valid over the fitting epochs.
However, as stated, the UVOT and ground-based data overlap
only marginally. We thus produced two SEDs. However, still
assuming the underlying spectrum remains constant, we could fit the
two SEDs simultaneously, sharing the intrinsic spectral slope and the
extinction, but allowing the normalization of the two SEDs to vary
independently.

The result is shown in Fig.~\ref{SEDFig}, and values are given in
Table~\ref{SEDFits}. A fit without dust results in a very red slope
and a bad fit. Fits with dust (based on \citealt{Pei1992ApJ}) improve
the fit significantly, but the shape of the double SED is not
simple. The usual dust that fits GRB afterglows well, that of the
Small Magellanic Cloud (SMC), also manages to fit the ground-based SED quite
well, but it fails to fit the UVOT SED, strongly underestimating the
UV emission. The dust of the Large Magellanic Cloud (LMC)
represents a strong improvement, but it results in a negative spectral
slope, which is very unlikely. The best fit is given by using Milky Way
(MW) dust with a strong 2175~\AA\ bump but shallow UV slope, only this
dust type is capable of modeling the whole double SED
adequately. Intriguingly, the derived values are very similar to
those from the SMC dust fit (Table~\ref{SEDFits}), but
the MW model agrees fully with the data. An upturn in the bluest
region of the X-shooter spectrum is also reported by \citet[][]{Fynbo2014AA} as would be expected from a MW
dust extinction model (however, see
  \citealt{Heintz2017AA}). Such a high extinction value is similar to
several other sightlines featuring a 2175~\AA\ bump, such as those
toward GRB 070802
\citep{Kruehler2008ApJ,Eliasdottir2009ApJ,Kann2010ApJ,Zafar2012ApJ},
GRB 120119A \citep[][Kann et al., in prep.]{Morgan2014MNRAS}, GRB
180325A \citep[][Kann et al., in prep.]{Zafar2018ApJ}, and GRB
190114C (Th\"one et al., in prep.).

\cite{Fynbo2014AA} reported an intrinsic spectral slope of the late
(post-prompt emission flares) X-ray spectrum of
$\beta_{\rm X}=0.75\pm0.07$. This is a hard spectrum already, but assuming a
cooling break between X-rays and optical implies $\beta_{\rm opt}=0.25$,
which is in excellent agreement with our result. \cite{Fynbo2014AA} also verified that
the cooling break lies between the X-ray and optical band. For such a
spectrum and the parametrization of \cite{Fitzpatrick2007ApJ},
\cite{Fynbo2014AA} found $A_{\rm V}=0.9$ mag, which is close to our result.
Furthermore, they reported that their spectrum cannot be fit with an MW extinction
law; however, they simply assumed $A_{\rm V}=0.8$ mag, which is significantly lower
than the value we derive. We independently analyzed the X-ray-to-optical SED and confirm the X-ray results of \cite{Fynbo2014AA}, with
a cooling break lying in the extreme UV. A fit purely with X-ray and
UVOT data also finds results in agreement with the UV-optical-NIR-only
fits within errors.

\cite{Heintz2017AA} presented strongly binned spectra from the first two
X-shooter epochs \citep{Fynbo2014AA}. They found that a subtraction of
their X-shooter host-galaxy spectrum (taken over a year after the GRB,
so there is no longer any SN contribution) essentially removes all flux
at wavelengths $<4000$~\AA; that is, the upturn seen in the X-shooter
afterglow spectra at bluest wavelengths was mostly pure host
contribution. They used this as evidence to rule out the ``extreme 2175
{\AA} bump'' model and derive a dust extinction model with an
extremely strong UV extinction. However, this is in contrast with our
clear UVOT $u$ detection and our UVOT-based SED, which shows a
relatively flat $b-u$ color that agrees with the existence of a 2175
{\AA} bump, as well as the lower S/N detection in the UVOT lenticular
filters, which should all be extremely damped if the
\cite{Heintz2017AA} dust model is correct. Additionally, from the
GROND SED, we do not detect a strong downturn yet in the
observer-frame $r^\prime$ band, bluer than the $<8000$~\AA\ downturn
\cite{Fynbo2014AA} and \cite{Heintz2017AA} derive from their X-shooter
spectra. We do not readily have a solution for this conundrum.

Independent of whether the extinction curve is an extreme 2175~\AA\
bump model as suggested by \cite{Fynbo2014AA} or the extreme UV
extinction model presented by \cite{Heintz2017AA}, the SN results we
report in Sect.~\ref{SNcomp}
are robust. The extinction \cite{Fynbo2014AA} found is similar to our
result, and the model of \cite{Heintz2017AA} also leads to a strong
correction for extinction, they find $A_{\rm V}=1.04$ mag, which is even closer to
our result. Furthermore, for the SN, only the $g^\prime r^\prime
i^\prime z^\prime$ data are important, and the two extinction models
are similar in this observer-frame wavelength range.

From our SED fit, we derive host-galaxy extinction correction
factors of $F_{\rm corr,g^\prime}=14.1\pm2.2$, $F_{\rm
  corr,r^\prime}=5.6\pm0.9$, $F_{\rm corr,i^\prime}=4.1\pm0.6$, and
$F_{\rm corr,z^\prime}=3.3\pm0.5$. Using these corrections and
applying the standard procedure for error propagation, we
achieve the final values given in the last column of
Table~\ref{SNFits}.

\begin{figure}[t!]
\centering 
\includegraphics[width=\columnwidth]{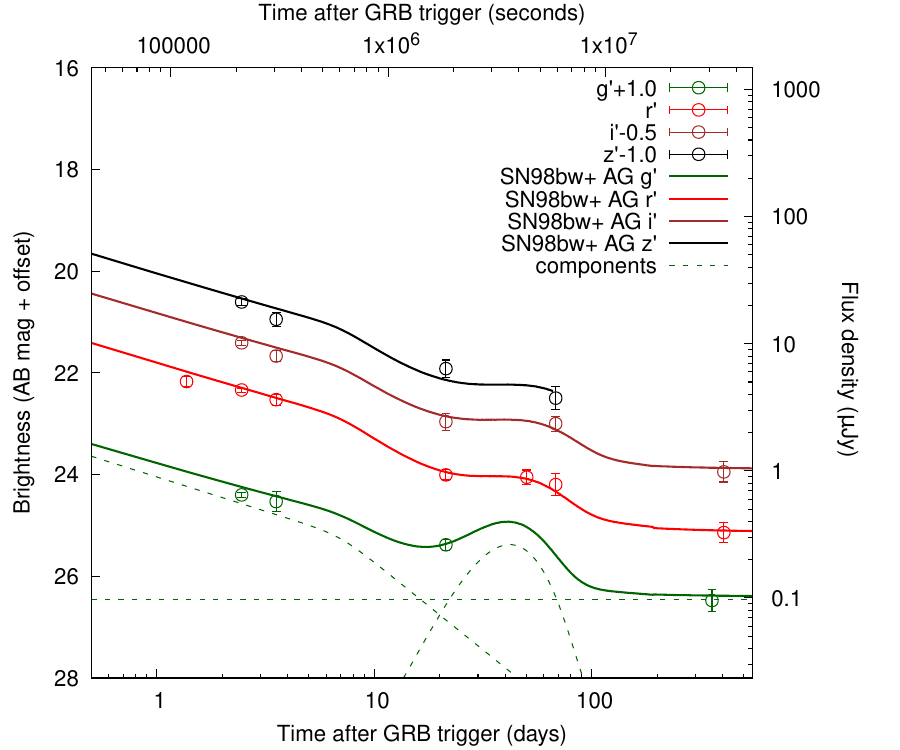}
\caption{ Late $g^\prime r^\prime i^\prime z^\prime$
      light curves of OT. For the $g^\prime$ band, we also show its
      breakdown into afterglow, host-galaxy, and SN flux. Data are
    corrected for Galactic extinction, given in the AB magnitude
    system, and additionally offset by the values given in the legend,
    but they are not corrected for the extinction in the host, as in
    Fig.~\ref{lc}. For the SN component, we used the best-fit values
    provided in Table~\ref{SNFits}.}
\label{lccompfig}
\end{figure}

\section{Discussion \label{discussion} }

Tables~\ref{SNFits} and \ref{altSNFits} show that despite the relatively large
difference in the chosen break time, the derived values are very
similar to each other, and the break time (within reasonable limits) has
only a minor influence on the SN. The second aspect is that the
resulting values are large. Generally, detecting an SN component at a
redshift of $\approx0.9$ is not expected and needs either a bright SN
or a very faint host galaxy. The contrast between the SN component and
the moderately bright host is immediately visible
\citep{Heintz2017AA}, implying a luminous SN.

\subsection{Light-curve break time \label{sec:break}}


 In Sect.~\ref{AG} we show how we found that a break in the optical
  light curve must exist between about three and 20 days; otherwise, at 21
  days the fit strongly overestimates the observed flux. Motivated by
  the observed shape of the \swift/XRT light curve, we adopted a break
  time of 7~d.
To study how the choice of the break time affects the evidence
  for an underlying SN component, we also modeled the optical light curve with a break at 12 d.  These two choices lead to two very different
results for the post-break decay slope: $\alpha_3=1.38\pm0.22$
for a break at seven days, and $\alpha_3=2.25\pm0.42$ for a break at 12 days.  The
resulting fits are indistinguishable in quality, and indeed, for
both cases we obtain $\chi^2/$d.o.f.=0.76 (Fig.~\ref{lc}).\footnote{These
  fits only use the $g^\prime$ data point at 21 d and not the upper
  limit at 68 d. However, the upper limit is not in disagreement with
  the SN 1998bw light curve shifted to $z$=0.889
  (Fig.~\ref{lc}).} The SN component is essentially unaffected in
  shape by the choice of the break time (but not in intrinsic
  luminosity).  Of course, these two choices do not map out the
  complete possible parameter region. For example, a break at 7 d (or
  even earlier) followed by a steep decay $>2$ is also possible, which
  would make the SN even more luminous. A late break with a shallow
  post-break decay would be in disagreement with the data at 21 days,
  however.

In conclusion, a later break time results in a more luminous
  SN. However, the size of this effect is small
  (Table~\ref{altSNFits}). Therefore, given that the derived
  luminosity of the SN does only marginally depend on the adopted
  break time, in the following we continue assuming a break at 7~d.
Finally, we note that the pre-break decay slope $\alpha_2$
was not influenced by our choice in break time, and the SEDs derived
from these fits were only marginally influenced by the break-time
choice.

\subsection{The SN luminosity in the $g^\prime$ band \label{sec:g} }

The SN light curve is best defined in the GROND $r^\prime$ band
  (3 data points), while less in $i^\prime$ and $z^\prime$ (2 data
  points each). In $g^\prime$, it relies on only one data point (at 21 days), but this is a bright detection when compared to the expected flux of SN 1998bw at that redshift as shown by the extreme $k_{g^\prime}$-value. This result is remarkable, and we must note several caveats. First at all, we must note that this extreme value needs to be seen relative to SN 1998bw, which shows a strong UV damping. 
  For example, this suppression of flux (due to the suppression of flux by metal line blanketing) was
not seen in SN 2011kl \citep{Greiner2015Nat,Mazzali2016MNRAS}.
  A second point is that
GRB/SN 140506A occurred at such a high redshift that the
wavelength region in the $g'$ band is no longer covered by the
redshifted SN 1998bw $UBVRI$ data set.  In other words, in order to predict
the $g'$-band light curve, we have to extrapolate to frequencies below
the $U$ band, which is very vague. The numerical procedure we had
originally developed in \cite{Zeh2004ApJ} assumes that for $\nu >
\nu(\rm U~band)$ the SN flux scales $\propto \nu^{-3}$,
normalized to the $U$-band flux. The results of this approach are
given in Tables~\ref{SNFits} and \ref{altSNFits}.  In
\cite{Klose2019AA}, where we present GRB-SN data of four events
between redshifts 0.4 and 0.8, it became clear, however, that this
procedure potentially under-predicts the flux of a GRB-SN in this
wavelength regime.

In order to improve our numerical procedure and overcome the above caveats, we modified our SN 1998bw
model as follows: we adopted a pure black body radiation of the SN
shell and calculated its (time-dependent) effective temperature
$T_{\rm eff}$ based on the observed $UBVRI$ broadband photometry as
it was published in \cite{Clocchiatti2011AJ}.  Based on this model,
with $T_{\rm eff}$ being a function of time, we extrapolated to
the blueshifted $g'$-band frequencies. This approach increases
the ``predicted'' peak brightness of SN 140506A by about 0.25 mag and
makes the $k_{g'}$ value correspondingly slightly smaller (a factor $\sim1.3$). 

A final consideration 
is that the $g^\prime$ value is based on a single data
point. We can muse as to how the luminosity would change if the
transient had also been detected at 68 days in $g^\prime$.  The upper
limit measured is significantly bluer than even the host-subtracted
$g^\prime-r^\prime$ color measured at 21 d. We wanted to know what the luminosity
would be if the SN were detected in $g^\prime$ at a fainter magnitude.  As
we have no real grasp of the color evolution, we assumed the
$g^\prime-r^\prime$ remains constant, and we derived a $g^\prime$
magnitude at 68 d based on the host-subtracted $r'$ detection and the
color measured at 21 d. We then added the host-galaxy flux to the
value, which ended up being 0.1 mag fainter in total than the upper
limit, and redid the fit (with a break time fixed to 7 d) with the
different host-galaxy magnitudes included, as before. Results for this
fit are shown in the last block of Table~\ref{altSNFits}. Hence, the SN
has become fainter in $g^\prime$ (as expected), but actually slightly
more luminous in the other bands; the fit is essentially unchanged in
quality, $\chi^2$/d.o.f. = 0.77. The main result, however, is that the
SN in $g^\prime$ is not much fainter than it was before (only 5\%).
Overall, although we consider the detection of an additional $g^\prime$-component to the late-time afterglow to be real, it remains a matter of speculation how luminous the SN associated with GRB 140506A was in the blueshifted $g^\prime$ band.

\subsection{Color evolution as evidence of an emerging SN component}

\begin{figure}[t]
\begin{center}
\centering 
\includegraphics[width=\columnwidth]{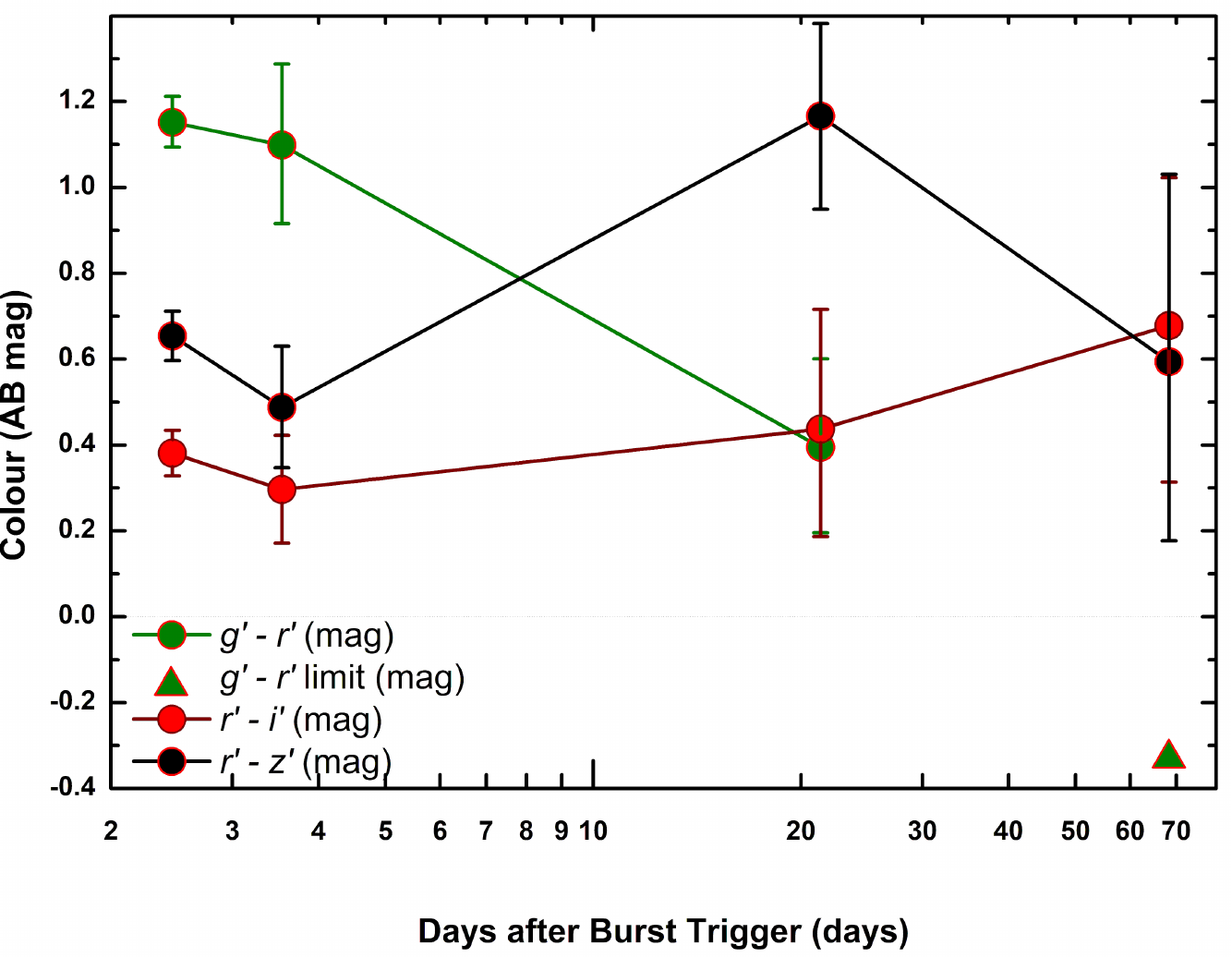}
\caption{Color evolution of optical transient following GRB
  140506A. Data are corrected for Galactic extinction,
  host-galaxy-subtracted, and given in AB magnitudes. The first two
  epochs evolve achromatically within errors, but then a clear color
  change is seen after three weeks, with the transient becoming both
  bluer in $g^\prime-r^\prime$ and redder in
  $r^\prime-z^\prime$. There is likely further evolution in the last
  epoch at 68 d, but the transient is no longer detected in $g^\prime$ to a non-constraining limit, and the errors are large in the two
  other colors. }
\label{color}
\end{center}
\end{figure}

Between four and 21 days, the broadband SED of the OT changed,
$g^\prime-r^\prime$ was decreasing, and $r^\prime-z^\prime$ was
increasing (Fig.~\ref{color}).   
While a color evolution during the afterglow phase is not an unknown
phenomenon, it is relatively rare (e.g., GRB 091127
\citealt{Filgas2011AA}, GRB 111209A \citealt{Kann2018AA}, and GRB
130427A \citealt{Perley2014ApJ}) and has never been observed due to
pure afterglow light weeks after a burst (after correcting for an
emerging host-galaxy flux). It is best understood as being due to a
rising thermal component \citep[e.g.][]{Olivares2015AA}, and thus it is a strong piece of evidence
that an observed bump in an afterglow light curve represents an
underlying SN component. In fact, such color evolution as evidence of
an upcoming SN was already pointed out in the case of the very first
cosmologically remote GRB-SN (030329/2003dh;
\citealt{Zeh2003GCN..2081....1Z}).

The $r^\prime-z^\prime$ color of the OT (roughly corresponding
  to an $U$--$g^\prime$ color in the GRB rest frame) increases up to
1 mag at 21 days (i.e., at 11.6 days rest-frame time).
Also, the $g^\prime-r^\prime$ color evolution (roughly corresponding to the rest-frame $uvw1 - u$) at 21 days supports a similar evolution, though this is based on a single detection.
Such colors of
$U$--$g^\prime \sim$1--2 mags have already been observed between ten
and 20 days in other GRB-SNe including GRB-SN 120422A-SN2012bz,
130427A-SN2013cq, and 171205A-SN2017iuk (e.g.,
\citealt{Schulze2014AA,Becerra2017ApJ,Izzo2019Nature}) and GRB-SN
081007-2008hw, 091127-2009nz, 101219B-2010ma
(\citealt{Olivares2015AA}) and GRB 130831A (associated with a
bump-only SN; \citealt{Klose2019AA}). The last three events were all at a
redshift $z\sim$0.5, and thus their $g^\prime-i^\prime$ color
corresponded to the $r^\prime-z^\prime$ of GRB 140506A.  The OT
that followed GRB 111209A showed a similar color of 1~mag at
$\sim$1--13 days (rest-frame time) when we consider its absolute
magnitudes at 2735 and 4556~\AA~ (Table~2 in \citealt{Kann2019AA}).
In other words, the observed color evolution of the OT
  following GRB 140506A is a strong indicator of an emerging SN
  component.

We note that the observed color change could suggest a late-time contribution from the interaction of the SN ejecta with circumstellar material (CSM). CSM interaction can manifest itself through   discrete emission lines, as seen in SNe IIn \citep{Smith2017a}, but   also through a blue pseudo-continuum, similar to that seen in SNe   Ia-CSM, Ibn, Icn, and some SNe IIn \citep[e.g.,][]{Silverman2013a,     Hosseinzadeh2017a, Gal-Yam2022a, Perley2022a}. There is also a small but growing number of SNe Ic-BL with evidence of late-time  CSM interaction, such as SN 2017ens \citep{Chen2018ApJL} and SN     2023xxf \citep{Kuncarayakti2023a}. 
To further explore this color-change in GRB-SNe, we need to collect more observations in the rest-frame UV. This can be done by observing rare nearby events in the UV, or with deep optical observations of the more distant GRB-SNe.
  
\begin{table*}[t]
\caption{Alternative solution for GRB 140506A supernova component.}
\label{altSNFits} 
\centering                        
\begin{tabular}{l c c l}
\hline\hline   
Fit model & $s$ & $k$ (fitted) & $k$ (extinction-corrected) \\
\hline
12 day break & $1.97\pm0.11$ & $k_{g^\prime}=5.69\pm1.17$ & $k_{g^\prime}=80.45\pm20.80$ \\
 &  & $k_{r^\prime}=1.42\pm0.34$ & $k_{r^\prime}=7.92\pm2.27$ \\
 &  & $k_{i^\prime}=1.35\pm0.37$ & $k_{i^\prime}=5.58\pm1.75$ \\
 &  & $k_{z^\prime}=1.12\pm0.39$ & $k_{z^\prime}=3.72\pm1.40$ \\
 \hline
7 day break & $1.91\pm0.12$ & $k_{g^\prime}=5.17\pm1.23$ & $k_{g^\prime}=73.11\pm20.80$ \\
Two $g^\prime$ points &  & $k_{r^\prime}=1.23\pm0.40$ & $k_{r^\prime}=6.86\pm2.50$ \\
 &  & $k_{i^\prime}=1.25\pm0.41$ & $k_{i^\prime}=5.14\pm1.88$ \\
 &  & $k_{z^\prime}=1.04\pm0.42$ & $k_{z^\prime}=3.46\pm1.49$ \\
\hline \hline
\end{tabular}
\end{table*}

\subsection{The missing spectroscopic confirmation of the SN}

Spectra of the optical transient following GRB 140506A were obtained
by \cite{Fynbo2014AA} at 8.8~h, 33~h, and 52~d post burst. They call the
spectrum taken at 52 days a host-galaxy spectrum. However, our data
show that it was actually taken during the SN phase (which was not
recognized by these authors). This spectrum covers the wavelength
range of 660 -- 977~nm; that is, it is missing the blue part that might
show SN features (which, e.g., were covered in the X-shooter
spectrum of GRB 111209A-SN 2011kl;
\citealt{Greiner2015Nat,Mazzali2016MNRAS}). Furthermore, the spectrum
is of low S/N, and only a host-galaxy emission line is
detected. Therefore, while it is a spectrum taken during the SN, it
cannot count as an actual SN spectrum. Consequently, the SN following GRB 140506A is bump-only, with only category D evidence following the characterization of
\cite{Hjorth2012Book}.\footnote{A bump, but the inferred SN
properties are not fully consistent with other GRB-SNe, or the bump
was not well sampled, or there is no spectroscopic redshift of the
GRB. \citep{Hjorth2012Book}}

\subsection{GRB-SN 140506A compared with GRB 111209A-SN 2011kl} 

The luminosity factors of the supernova shown in the right part
  of Table~\ref{SNFits} are extreme, and significantly larger than
  those of even SN 2011kl, the SN associated with GRB 111209A
  \citep{Kann2019AA}.  We note that this GRB had a somewhat lower
redshift ($z=0.67702$, \citealt{Kann2018AA}), and therefore the
observer-frame bands (which are identical) are not the same in the
rest frame compared to GRB 140506A. The observer-frame $z^\prime$ band
of GRB 140506A roughly agrees with the observer-frame $i^\prime$ band
for GRB 111209A, both being close to the rest-frame $B-g^\prime$ band
(where the SN 1998bw template is also well defined). In this case,
it is $k_{z^\prime}^{\textnormal{GRB-SN 140506A}}=3.32\pm1.45$ versus
$k_{i^\prime}^{\textnormal{SN 2011kl}}=1.81\pm0.22$. Therefore, the SN
associated with GRB 140506A is the most luminous detected so far.
However, the error bars are large, and the difference is only slightly
above $1\sigma$.  In the case of GRB 140506A, the extinction is
clearly very large \citep[][and
  Sect.~\ref{SED}]{Fynbo2014AA,Heintz2017AA}, which is in contrast with the
(relatively small) extinction measured from the GRB 111209A afterglow
\citep{Kann2018AA,Kann2019AA}.

We also note that the X-ray light curve shows a deviation from
  a pure single power-law decay downward at about 6-7 d, that
  we explain as the possible jet break. However, the flux has
recovered by 14 d, and the further decay is in agreement with the
earlier afterglow decay. Such an evolution is very similar to the atypical X-ray light curve of GRB 111209A. \cite{Kann2019AA} speculated this might be a jet
break (which is detected in the UV-optical for GRB 111209A) combined
with rising X-ray emission associated with SN 2011kl. A similar
phenomenon may be visible here, associated with the extremely luminous
and blue SN. However, the sparse data mean this remains speculation.

\subsection{The GRB-SN luminosity-stretch diagram}

\begin{figure}[t]
\begin{center}
\centering 
\includegraphics[width=\columnwidth]{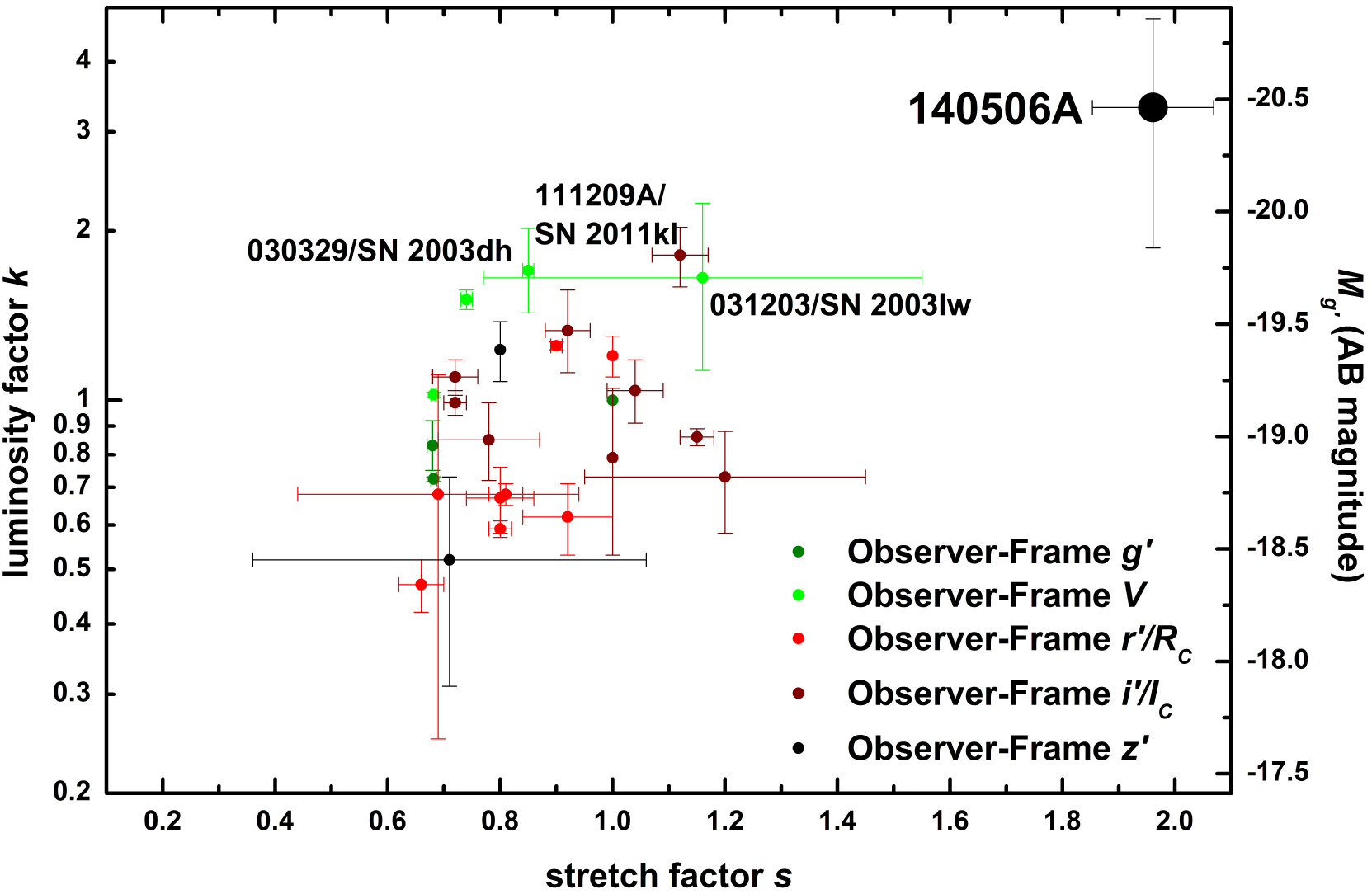}
\caption{Luminosity of GRB-SNe (given in luminosity versus the
  luminosity of an SN 1998bw template, $k$) versus their peak
    time (given in stretch versus the light curve of an SN 1998bw
  template, $s$). SN 1998bw is \emph{\emph{per definitionem}} at
  ($k,s=1,1$). These ($k,s$) values are derived from observer-frame
  filter fits given in the legend (and color-coded), which are close to
  the rest-frame $g^\prime$ filter; hence, they are all roughly
  comparable. The right-hand scale in absolute magnitude is therefore
  not completely precise. We highlight several luminous GRB-SNe. The
  SN bump associated with GRB 140506A stands out strongly but
  represents an extension of the luminosity-duration correlation that
  has been discussed in the literature 
    \citep[e.g.,][]{Schulze2014AA, Cano2017AdAst,Klose2019AA}.}
\label{ks}
\end{center}
\end{figure}

There has been discussion on the use of GRB-SNe as standard candles,
with an initial study done by \citet[][their figure
  11]{Schulze2014AA}, and expanded upon by
\cite{Cano2014ApJ,Cano2014MNRAS,Li2014AA1,Li2014ApJ}. To place the SN
of GRB 140506A into such a context, we collected data from the samples
of \citet[][]{Ferrero2006AA}, \citet[][]{Thoene2011Nature}, and \citet[][]{Klose2019AA} and use, similarly
to \cite{Schulze2014AA}, a ``quasi rest-frame $g^\prime$ band'', using
($k,s$) values where the observer-frame band corresponds reasonably to
the rest-frame $g^\prime$ band. This is an imprecise process, 
but it is sufficient to yield a qualitative result. We show the resulting
data in Fig.~\ref{ks}. We color-code the observer-frame bands in
which the SNe were fit, which acts as a rough redshift measure.

Figure~\ref{ks} shows that the GRB 140506A supernova is a
strong outlier compared to the other GRB-SNe, even GRB
111209A-SN~2011kl, both being far more luminous and far slower in
terms of light-curve evolution. However, the combination of these two
extremes places the GRB 140506A-SN on the extension of the GRB-SN
Phillips relation first discussed by \cite{Schulze2014AA}. A more
quantitative analysis will be performed in future work. In
Fig.~\ref{ks}, we also show the absolute magnitude in $g^\prime$ on
the right y axis. We took the $M_B$ of SN 1998bw, corrected it for
host-galaxy extinction \citep{Kann2019AA}, and estimated
$B-g^\prime\,(\textrm{SN 1998bw})=0.2$ mag. We note the data points do
not agree with this scale exactly, as they are only close to the
rest-frame $g^\prime$ band.

\subsection{Dust destruction and color evolution}

\cite{Fynbo2014AA} reported on a small but definite shift in the
location of the steep flux depression from their first to their second
spectrum. We cannot check for a respective color change as there is
no multicolor photometry during the first epoch. \cite{Heintz2017AA}
pointed out that this effect may be linked to the emission region in the
second epoch being significantly larger. The very high collimation of
GRB afterglows at very early times implies that if dust destruction
were to be induced in a dust curtain near the GRB, the affected area
would be significantly smaller than the emission region at later
times. \cite{Morgan2014MNRAS} reported the signature of dust destruction
in the very early afterglow of GRB 120119A; however, even after the
extinction becomes constant, it is still very high, $A_{\rm V}\approx1.1$
mag \citep[][Kann et al., in prep.]{Morgan2014MNRAS,Japelj2015A&A,Zafar2018MNRAS}, similar to our
result for GRB 140506A. The color evolution induced by dust burning
is only detected in the first $\approx100$ rest-frame seconds, however, in
the case of GRB 120119A. At such times, the only data we have for GRB
140506A are the unfiltered UVOT $white$ data beginning 56 rest-frame s
post-trigger (though, we note the existence of several further strong
X-ray flares in Fig.~\ref{lc}). While these few data points do not show
a flatter evolution or even rise (emergence from a dusty envelope was
a model suggested for the light-curve behavior of GRB 030418,
\citealt{Rykoff2004ApJ}), the lack of any color evolution does not allow us
to set any constraints on the potential existence of dust burning, or
lack thereof.

\section{Conclusions \label{conclusions}}

We studied the light curve, afterglow SED, and potential SN of
the ordinary GRB 140506A, which so far had only been remarkable for its
high line-of-sight extinction, peculiar dust law, and spectral
features \citep{Fynbo2014AA,Heintz2017AA}.
We find that despite clear strong reddening, the afterglow is detected
all the way into the UV at early times, allowing us to create a
broadband SED that confirms the high extinction and favors an MW-type
extinction law with a clear 2175~\AA\ bump.

The late-time transient shows a clear color change and a long plateau
phase before reaching a fainter host-galaxy level, which we interpret
as the SN following GRB 140506A. This is a remarkable result,
especially the strong detection in the observer-frame $g^\prime$ band,
considering the high redshift of $z\approx0.9$ (the highest redshift
at which GRB-SNe have been detected so far is $z\approx1$,
\citealt{DellaValle2003AA,Masetti2005AA}). It becomes even more
remarkable once the strong extinction is corrected for, resulting in
the most luminous (but also most slowly evolving) GRB-SN detected so
far.  
The SN remained luminous deep into the UV, similar to GRB 111209A-SN
2011kl. The latter event was, however, associated with a remarkable,
ultra-long GRB
\citep{Gendre2013ApJ,Greiner2015Nat,Mazzali2016MNRAS,Kann2018AA,Kann2019AA},
whereas GRB 140506A is unremarkable in terms of both duration and
energetics.

The very luminous but also very slowly evolving SN of GRB 140506A
hints that a luminosity-peak time correlation for GRB-SNe, which would
have implications for cosmological measurements, exists and extends
all the way into a part of the parameter space that has so far not
been populated. It makes it clear that late-time follow-up of GRB
afterglows at $z\lesssim1$ can always yield interesting surprises
(unless the host galaxy is too bright and masks the SN emission, which
can happen even at significantly lower redshifts as in the case of
the faint SN accompanying XRF 100418A,
\citealt{Niino2012PASJ,deUgartePostigo2018AA2}). More overly luminous
SNe such as the one associated with GRB 140506A need to be found to allow
the use of GRB-SNe as robust cosmological tracers.


\begin{acknowledgements}

The co-authors wish to thank the editor for allowing us to complete this study after the recent loss
of our colleague David Alexander Kann, R.I.P. 10. 03. 2023. We thank the referee for the very valuable report which helped improve the clarity of the paper. We also thank S. Schmidl for help with the figures.
DAK acknowledges support from Spanish National Research Project
RTI2018-098104-J-I00 (GRBPhot). AdUP and CCT acknowledge support from
Ram\'on y Cajal fellowships RyC-2012-09975 and RyC-2012-09984 and the
Spanish Ministry of Economy and Competitiveness through projects
AYA2014-58381-P and AYA2017-89384-P, AdUP furthermore from the BBVA
foundation. 
AR acknowledge support from PRIN-MIUR 2017 (grant 20179ZF5KS). SK acknowledges support by the
Th\"uringer Ministerium f\"ur Bildung, Wissenschaft und Kultur under
FKZ 12010-514 and by grants DFG Kl 766/16-1 and 766/16-3. SS
is supported by LBNL Subcontract NO. 7707915. This work
made use of data supplied by the UK Swift Science Data Centre at the
University of Leicester.

\end{acknowledgements}

\bibliographystyle{aa_url}
\bibliography{mybib} 

\clearpage
\onecolumn

\begin{appendix}

\section{Additional observational data}\label{append1}


\begin{longtable}{l c l}
\caption{UVOT observations of GRB 140506A.}\\
\label{obslog} 
$\Delta$t (days) & mag & filter \\
\endfirsthead
\caption{continued.}\\
\hline
$\Delta$t (days) & mag & filter \\
\hline  
\endhead
\hline
\endfoot
\vspace{1mm}
0.007419        & $     21.334  ^{+     2.296   }_{-    0.685   }$ &    $uvw2$  \\\vspace{1mm}
0.009423        & $     21.434  ^{+     4.156   }_{-    0.741   }$ &    $uvw2$  \\\vspace{1mm}
0.012113        & $     21.153  ^{+     2.116   }_{-    0.672   }$ &    $uvw2$  \\\vspace{1mm}
0.014116        & $     21.462  ^{+     4.645   }_{-    0.745   }$ &    $uvw2$  \\\vspace{1mm}
0.073942        & $     22.928  ^{+     1.628   }_{-    0.624   }$ &    $uvw2$  \\\vspace{1mm}
0.016118        & $ >   20.447                                  $ &     $uvw2$ UL      \\\vspace{1mm}
0.018121        & $ >   20.991                                  $ &     $uvw2$ UL      \\\vspace{1mm}
0.195006        & $ >   23.420                                  $ &     $uvw2$ UL      \\\vspace{1mm}
0.535097        & $ >   23.400                                  $ &     $uvw2$ UL      \\\vspace{1mm}
0.795653        & $ >   22.917                                  $ &     $uvw2$ UL      \\\vspace{1mm}
5.834459        & $ >   24.823                                  $ &     $uvw2$ UL      \\\vspace{1mm}
13.803396       & $ >   23.589                                  $ &     $uvw2$ UL      \\\vspace{1mm}
17.937837       & $ >   23.488                                  $ &     $uvw2$ UL      \\ \vspace{1mm}
0.007986        & $     20.189  ^{+     0.912   }_{-    0.489   }$ &    $uvm2$  \\\vspace{1mm}
0.012679        & $     20.731  ^{+     1.594   }_{-    0.620   }$ &    $uvm2$  \\\vspace{1mm}
0.078682        & $     21.707  ^{+     0.623   }_{-    0.393   }$ &    $uvm2$  \\\vspace{1mm}
0.009999        & $ >   19.788                                  $ &     $uvm2$ UL      \\\vspace{1mm}
0.014680        & $ >   20.153                                  $ &     $uvm2$ UL      \\\vspace{1mm}
0.016684        & $ >   19.684                                  $ &     $uvm2$ UL      \\\vspace{1mm}
0.062059        & $ >   21.734                                  $ &     $uvm2$ UL      \\\vspace{1mm}
0.340650        & $ >   22.883                                  $ &     $uvm2$ UL      \\\vspace{1mm}
2.361840        & $ >   23.194                                  $ &     $uvm2$ UL      \\\vspace{1mm}
2.529452        & $ >   22.703                                  $ &     $uvm2$ UL      \\\vspace{1mm}
6.797721        & $ >   22.957                                  $ &     $uvm2$ UL      \\\vspace{1mm}
10.725055       & $ >   23.950                                  $ &     $uvm2$ UL      \\\vspace{1mm}
18.395811       & $ >   24.251                                  $ &     $uvm2$ UL      \\\vspace{1mm}
22.595256       & $ >   23.429                                  $ &     $uvm2$ UL      \\\vspace{1mm}
26.761876       & $ >   23.785                                  $ &     $uvm2$ UL      \\\vspace{1mm}
30.580328       & $ >   24.084                                  $ &     $uvm2$ UL      \\\vspace{1mm}
0.008277        & $     19.318  ^{+     0.468   }_{-    0.326   }$ &    $uvw1$  \\\vspace{1mm}
0.012966        & $     20.678  ^{+     1.647   }_{-    0.626   }$ &    $uvw1$  \\\vspace{1mm}
0.016978        & $     20.198  ^{+     0.994   }_{-    0.510   }$ &    $uvw1$  \\\vspace{1mm}
0.064437        & $     22.190  ^{+     0.862   }_{-    0.474   }$ &    $uvw1$  \\\vspace{1mm}
0.081058        & $     22.188  ^{+     1.442   }_{-    0.598   }$ &    $uvw1$  \\\vspace{1mm}
0.014971        & $ >   20.073                                  $ &     $uvw1$ UL      \\\vspace{1mm}
0.348816        & $ >   21.957                                  $ &     $uvw1$ UL      \\\vspace{1mm}
0.465948        & $ >   22.813                                  $ &     $uvw1$ UL      \\\vspace{1mm}
0.734638        & $ >   23.819                                  $ &     $uvw1$ UL      \\\vspace{1mm}
3.201317        & $ >   23.158                                  $ &     $uvw1$ UL      \\\vspace{1mm}
7.797540        & $ >   23.376                                  $ &     $uvw1$ UL      \\\vspace{1mm}
11.663524       & $ >   24.137                                  $ &     $uvw1$ UL      \\\vspace{1mm}
15.791810       & $ >   24.423                                  $ &     $uvw1$ UL      \\\vspace{1mm}
0.003941        & $     17.923  ^{+     0.131   }_{-    0.117   }$ &    $u$     \\\vspace{1mm}
0.004288        & $     17.838  ^{+     0.124   }_{-    0.111   }$ &    $u$     \\\vspace{1mm}
0.004691        & $     18.555  ^{+     0.166   }_{-    0.144   }$ &    $u$     \\\vspace{1mm}
0.005154        & $     18.484  ^{+     0.160   }_{-    0.140   }$ &    $u$     \\\vspace{1mm}
0.005618        & $     18.884  ^{+     0.212   }_{-    0.177   }$ &    $u$     \\\vspace{1mm}
0.006137        & $     18.763  ^{+     0.178   }_{-    0.153   }$ &    $u$     \\\vspace{1mm}
0.006546        & $     18.657  ^{+     0.278   }_{-    0.221   }$ &    $u$     \\\vspace{1mm}
0.008560        & $     18.816  ^{+     0.352   }_{-    0.265   }$ &    $u$     \\\vspace{1mm}
0.013254        & $     19.328  ^{+     0.753   }_{-    0.440   }$ &    $u$     \\\vspace{1mm}
0.015102        & $     19.580  ^{+     0.303   }_{-    0.237   }$ &    $u$     \\\vspace{1mm}
0.017261        & $     19.820  ^{+     2.739   }_{-    0.708   }$ &    $u$     \\\vspace{1mm}
0.066808        & $     20.631  ^{+     0.260   }_{-    0.210   }$ &    $u$     \\\vspace{1mm}
0.083428        & $ >   20.218                                  $ &     $u$ UL      \\\vspace{1mm}
0.400859        & $ >   24.178                                  $ &     $u$ UL      \\\vspace{1mm}
0.484796        & $ >   19.811                                  $ &     $u$ UL      \\\vspace{1mm}
0.743324        & $ >   21.479                                  $ &     $u$ UL      \\\vspace{1mm}
4.798999        & $ >   22.811                                  $ &     $u$ UL      \\\vspace{1mm}
8.223336        & $ >   23.195                                  $ &     $u$ UL      \\\vspace{1mm}
20.363598       & $ >   22.264                                  $ &     $u$ UL      \\\vspace{1mm}
24.395194       & $ >   23.851                                  $ &     $u$ UL      \\\vspace{1mm}
28.659326       & $ >   23.269                                  $ &     $u$ UL      \\\vspace{1mm}
0.006843        & $     19.346  ^{+     1.227   }_{-    0.561   }$ &    $b$     \\\vspace{1mm}
0.008847        & $     19.083  ^{+     1.174   }_{-    0.551   }$ &    $b$     \\\vspace{1mm}
0.015394        & $     19.763  ^{+     0.801   }_{-    0.456   }$ &    $b$     \\\vspace{1mm}
0.017547        & $     19.533  ^{+     2.869   }_{-    0.713   }$ &    $b$     \\\vspace{1mm}
0.069184        & $     20.504  ^{+     0.438   }_{-    0.311   }$ &    $b$     \\\vspace{1mm}
0.130621        & $     21.261  ^{+     0.370   }_{-    0.275   }$ &    $b$     \\\vspace{1mm}
0.013540        & $ >   18.433                                  $ &     $b$ UL      \\\vspace{1mm}
0.084803        & $ >   18.798                                  $ &     $b$ UL      \\\vspace{1mm}
0.411421        & $ >   22.854                                  $ &     $b$ UL      \\\vspace{1mm}
0.663690        & $ >   21.600                                  $ &     $b$ UL      \\\vspace{1mm}
0.001132        & $     17.034  ^{+     0.310   }_{-    0.241   }$ &    $v$     \\\vspace{1mm}
0.007704        & $     18.110  ^{+     0.536   }_{-    0.357   }$ &    $v$     \\\vspace{1mm}
0.012396        & $     17.866  ^{+     0.527   }_{-    0.353   }$ &    $v$     \\\vspace{1mm}
0.016400        & $     18.182  ^{+     0.803   }_{-    0.457   }$ &    $v$     \\\vspace{1mm}
0.059688        & $     20.106  ^{+     0.518   }_{-    0.349   }$ &    $v$     \\\vspace{1mm}
0.076311        & $     19.881  ^{+     0.543   }_{-    0.360   }$ &    $v$     \\\vspace{1mm}
0.009704        & $ >   18.423                                  $ &     $v$ UL      \\\vspace{1mm}
0.014398        & $ >   17.990                                  $ &     $v$ UL      \\\vspace{1mm}
0.330130        & $ >   20.637                                  $ &     $v$ UL      \\\vspace{1mm}
0.592875        & $ >   20.325                                  $ &     $v$ UL      \\\vspace{1mm}
0.802144        & $ >   20.476                                  $ &     $v$ UL      \\\vspace{1mm}
0.001360        & $     17.676  ^{+     0.084   }_{-    0.078   }$ &    $white$ \\\vspace{1mm}
0.001476        & $     17.928  ^{+     0.096   }_{-    0.088   }$ &    $white$ \\\vspace{1mm}
0.001591        & $     17.949  ^{+     0.097   }_{-    0.089   }$ &    $white$ \\\vspace{1mm}
0.001707        & $     17.943  ^{+     0.096   }_{-    0.088   }$ &    $white$ \\\vspace{1mm}
0.001823        & $     17.835  ^{+     0.090   }_{-    0.083   }$ &    $white$ \\\vspace{1mm}
0.001939        & $     18.115  ^{+     0.105   }_{-    0.096   }$ &    $white$ \\\vspace{1mm}
0.002055        & $     18.058  ^{+     0.101   }_{-    0.093   }$ &    $white$ \\\vspace{1mm}
0.002170        & $     18.115  ^{+     0.105   }_{-    0.096   }$ &    $white$ \\\vspace{1mm}
0.002286        & $     18.212  ^{+     0.110   }_{-    0.100   }$ &    $white$ \\\vspace{1mm}
0.002458        & $     18.279  ^{+     0.080   }_{-    0.075   }$ &    $white$ \\\vspace{1mm}
0.002689        & $     18.281  ^{+     0.080   }_{-    0.074   }$ &    $white$ \\\vspace{1mm}
0.002920        & $     18.311  ^{+     0.082   }_{-    0.076   }$ &    $white$ \\\vspace{1mm}
0.007126        & $     19.255  ^{+     0.160   }_{-    0.139   }$ &    $white$ \\\vspace{1mm}
0.009129        & $     19.735  ^{+     0.275   }_{-    0.219   }$ &    $white$ \\\vspace{1mm}
0.011030        & $     19.916  ^{+     0.115   }_{-    0.104   }$ &    $white$ \\\vspace{1mm}
0.013822        & $     19.986  ^{+     0.410   }_{-    0.297   }$ &    $white$ \\\vspace{1mm}
0.015823        & $     20.398  ^{+     0.720   }_{-    0.429   }$ &    $white$ \\\vspace{1mm}
0.017829        & $     20.103  ^{+     0.556   }_{-    0.366   }$ &    $white$ \\\vspace{1mm}
0.071559        & $     20.909  ^{+     0.160   }_{-    0.140   }$ &    $white$ \\\vspace{1mm}
0.141189        & $     22.429  ^{+     0.265   }_{-    0.213   }$ &    $white$ \\\vspace{1mm}
1.806576        & $     23.528  ^{+     0.717   }_{-    0.428   }$ &    $white$ \\\vspace{1mm}
0.417553        & $ >   21.972                                  $ &     $white$ UL      \\\vspace{1mm}
0.672448        & $ >   24.344                                  $ &     $white$ UL      \\\vspace{1mm}
0.966973        & $ >   24.512                                  $ &     $white$ UL      \\\vspace{1mm}
1.265749        & $ >   23.492                                  $ &     $white$ UL      \\\vspace{1mm}
1.468435        & $ >   24.223                                  $ &     $white$ UL      \\\vspace{1mm}
1.976269        & $ >   23.667                                  $ &     $white$ UL      \\ 
\end{longtable}
\tablefoot{All data are in AB magnitudes and not corrected for Galactic foreground extinction. Midtimes were derived with the geometric mean of start and stop times. $t=sqrt[(t_1-t_0)\times(t_2-t_0)]$, hereby $t_{1.2}$ are the absolute start and stop times, and $t_0$ is the \emph{Swift} trigger time. To obtain Vega magnitudes, it is $uvw2_{AB}-uvw2_{Vega}=1.73$ mag, $uvm2_{AB}-uvm2_{Vega}=1.69$ mag, $uvw1_{AB}-uvw1_{Vega}=1.51$ mag, $u_{AB}-u_{Vega}=1.02$ mag, $b_{AB}-b_{Vega}=-0.13$ mag, $v_{AB}-v_{Vega}=-0.01$ mag, and $white_{AB}-white_{Vega}=0.80$ mag (as given at http://swift.gsfc.nasa.gov/analysis/uvot\_digest/zeropts.html). Corrections for Galactic extinction are, using $E_{(B-V)}=0.080$ mag \citep{Schlafly2011ApJ} and the Galactic extinction curve of \cite{Cardelli1989ApJ}: $A_{uvw2}=0.728$ mag, $A_{uvm2}=0.763$ mag, $A_{uvw1}=0.539$ mag, $A_u=0.405$ mag, $A_b=0.328$ mag, $A_v=0.256$ mag, and $A_{white}=0.397$ mag.}


\end{appendix}

\end{document}